\documentclass[12pt]{article}

\usepackage{graphicx}
\usepackage{latexsym}
\usepackage{amscd}
\usepackage[cp1251]{inputenc}
\usepackage[english,russian]{babel}
\usepackage{amsmath,  eucal}
\usepackage{amssymb}
\usepackage{indentfirst}
\usepackage[small,centerlast]{caption2}
\usepackage{longtable}
\usepackage[dvipsnames]{color}% - это необходимо для правки
\usepackage{cite}
\graphicspath{{figure/}}

\makeatletter
\renewcommand{\@biblabel}[1]{#1.\hfill}

\renewcommand{\Re}{\mathop{\rm Re\,}}
\renewcommand{\Im}{\mathop{\rm Im\,}}
\makeatother {\renewcommand{\baselinestretch}{1.1}

\makeatletter
\renewcommand{\section}{\@startsection {section}{1}{\z@}%
                                   {-3.5ex \@plus -1ex \@minus -.2ex}%
                                   {2.3ex \@plus.2ex}%
                                   {\normalfont\Large\uppercase}}
\renewcommand{\subsection}{\@startsection{subsection}{2}{\z@}%
                                     {-3.25ex\@plus -1ex \@minus -.2ex}%
                                     {1.5ex \@plus .2ex}%
                                     {\normalfont\large\itshape}}
\renewcommand{\subsubsection}{\@startsection{subsubsection}{3}{1em}%
                                     {-3.25ex\@plus -1ex \@minus -.2ex}%
                                     {-1.5em \@plus .2em}%
                                     {\normalfont\normalsize\bfseries}}

\makeatother

\begin{document}

\newcommand{\mc}[1]{\mathcal{#1}}
\newcommand{\E}{\mc{E}}
\thispagestyle{empty} \large
\renewcommand{\abstractname}{}
\centerline{\textbf{\Large PlASMA DYNAMICS}}

 \begin{center}
{\Large \textbf{GENERATION OF THE LONGITUDINAL CURRENT BY THE
TRANSVERSAL ELECTROMAGNETIC FIELD IN CLASSICAL AND QUANTUM PLASMAS}}
\medskip \bigskip

{\large \textbf{A.V. Latyshev, A.A. Yushkanov}}
\medskip

\textit{Moscow State Regional University}
\\[10pt]

\end{center}

\begin{abstract}
\noindent
From Vlasov kinetic equation for collisionless plasmas
distribution  function in square-law approximation on size
of electromagnetic field is received. Formulas for calculation
electric current at any temperature (any degree of degeneration
of electronic gas) are deduced. The case of small values of the
wave  numbers is considered.
It is shown, that the nonlinearity account leads to occurrence
the longitudinal electric current directed along a wave vector.
This longitudinal current orthogonal to  known transversal classical current,
received at the linear analysis.
From the kinetic equation with Wigner integral for
collisionless quantum plasma distribution  function is received
in square-law on vector potential approximation.
Formulas for calculation electric current at any temperature are deduced.
The case of small values of wave number is considered. It is shown, that
size of a longitudinal current at small values of wave number and
for classical plasma and for quantum plasma coincide.
Graphic comparison of dimensionless size of a current
quantum and classical plasma is made.
\end{abstract}

\section*{Introduction}

Dielectric permeability of quantum plasma was studied by many
authors \cite{Klim} -- \cite{Dressel}.
It is one of the major
characteristics of plasma also it is applied in the diversified questions
physicists of plasma \cite{Dressel} -- \cite{Lat4}.

Let us notice, that in work \cite{Klim} the formula
for calculation of longitudinal dielectric permeability into
quantum plasma for the
first time has been deduced. Then the same formula has been deduced and in work
\cite{Lin}.

In the present work formulas for calculation electric
current in classical and quantum collisionless plasma
at any temperature, at the any
degrees of degeneration of electronic gas are deduced.

The approach developed by Klimontovich and Silin \cite{Klim}
is thus generalised.

At the decision of the kinetic equation we consider as in decomposition
of distribution function, and in decomposition and sizes
of the self-consistent
electromagnetic field and Wigner integral
the sizes proportional
to square of intensity or potential of an external electromagnetic field.

Electric current expression consists of two composed.
The first composed, linear on intensity of an electromagnetic field, is
known classical expression of an electric current.
This electric current is directed along the electromagnetic
fields. The second composed represents an electric current,
which is pro\-por\-tio\-nal to an intensity square of
electromagnetic field. The second current
it is perpendicular to the first and it is directed along the wave
vector. Occurrence of the second current comes to light the spent account
nonlinear character
interactions of an electromagnetic field with classical and quantum plasma.

In works \cite{Gins} and \cite{Zyt} nonlinear effects are studied
into plasma. In work \cite{Zyt} the nonlinear current was used, in
particulars, in probability questions disintegration processes. We will note,
that in work \cite{Zyt2} it is underlined existence
nonlinear current along a wave vector
(see the formula (2.9) from \cite {Zyt2}).

\section{Classical plasmas}

Let us show, that in case of the classical plasma described
by the Vlasov equation, the longitudinal current is generated
and we will calculate its density. On
existence of this current was specified more half a century ago
\cite {Zyt2}.
We take the Vlasov equation describing of behaviour
of collisionless plasmas
$$
\dfrac{\partial f}{\partial t}+\mathbf{v}\dfrac{\partial f}{\partial
\mathbf{r}}+
e\bigg(\mathbf{E}+
\dfrac{1}{c}[\mathbf{v},\mathbf{H}]\bigg)
\dfrac{\partial f}{\partial\mathbf{p}}=0.
\eqno{(1.1)}
$$

Electric and magnetic fields are connected with the vector
potential by equalities
$$
\mathbf{E}=-\dfrac{1}{c}\dfrac{\partial \mathbf{A}}{\partial
t}=\dfrac{i\omega}{c}\mathbf{A},\;\qquad
\mathbf{H}={\rm rot} \mathbf{A}.
$$

Therefore,
$$
{\bf H}=\dfrac{ck}{\omega}E_y\cdot(0,0,1),\qquad
{\bf [v,H}]=\dfrac{ck}{\omega}E_y\cdot (v_y,-v_x,0),
$$
$$
e\bigg(\mathbf{E}+\dfrac{1}{c}[\mathbf{v},\mathbf{H}]\bigg)
\dfrac{\partial f}{\partial\mathbf{p}}=
\dfrac{e}{\omega}E_y\Big[kv_y\dfrac{\partial f}{\partial p_x}+
(\omega-kv_x)\dfrac{\partial f}{\partial p_y}\Big].
$$

Let us operate with the successive-approximations method,
considering as small parametre size of intensity of electric field.
Let us copy the equation (1.1) in the form
$$
\dfrac{\partial f^{(k)}}{\partial t}+v_x\dfrac{\partial f^{(k)}}{\partial x}=
$$
$$
=-
\dfrac{eE_y}{\omega}\Bigg[kv_y\dfrac{\partial f^{(k-1)}}{\partial p_x}+
(\omega-kv_x)\dfrac{\partial f^{(k-1)}}{\partial p_y}\Bigg],\qquad k=1,2.
\eqno{(1.2)}
$$ \bigskip

Here $f^{(0)}=f_0(v)$ is the absolute Fermi---Dirac distribution,
$$
f_0(v)=\Big[1+\exp\dfrac{\E-\mu}{k_BT}\Big]^{-1},
$$
$\E=\dfrac{mv^2}{2}$ is the electrons energy, $\mu$ is the
chemical potential of electronic gas, $k_B$ is the Boltzmann
constant, $T$ is the plasma temperature.

Let us consider, that intensity of electric field varies
harmoniously:
$$
{\bf E} = {\bf E}_0e^{{\bf i (kr-\omega t)}}.
$$
Wave
vector we will direct along an axis $x $: $ {\bf k} =k (1,0,0) $, and
intensity of electric field we will direct along an axis $y $:
$ {\bf E} =E_y (0,1,0) $.

We notice that
$$
[\mathbf{v,H}]\dfrac{\partial f_0}{\partial \mathbf{p}}=0,
$$
because
$$
\dfrac{\partial f_0}{\partial \mathbf{p}}\sim \mathbf{v}.
$$

We search for the solution as a first approximation in the form
$$
f^{(1)}=f_0(P)+f_1,
$$
where $f_1\sim E_y$.

In this approximation the equation (1.2) becomes simpler
$$
\dfrac{\partial f_1}{\partial t}+v_x\dfrac{\partial f_1}{\partial x}=
-eE_y\dfrac{\partial f_0}{\partial p_y}.
\eqno{(1.3)}
$$

From (1.3) we receive
$$
f_1=\dfrac{2ieE_y}{p_T}\dfrac{P_y g(P)}{\omega-kv_TP_x},
\eqno{(1.4)}
$$
where
$$
g(P)=\dfrac{e^{P^2-\alpha}}{(1+e^{P^2-\alpha})^2},\qquad
{\bf P}=\dfrac{{\bf p}}{p_T}.
$$

Here $v_T=\sqrt{2k_BT/m}$ is the thermal electrons velocity,
$p_T=mv_T$ is the thermal electrons momentum, ${\bf P}$
is the dimensionless momentum.

In the second approximation for the solution of the equation
(1.2) we search in the form
$$
f^{(2)} =f^{(1)} +f_2,
$$
where $f_2\sim E_y^2$.

From the equation (1.2) it is found
$$
f_2=\dfrac{e^2E_y^2}{p_T^2\omega}
\Bigg[kv_TP_y^2\dfrac{\partial}{\partial P_x}
\Big(\dfrac{g(P)}{\omega-kv_TP_x}\Big)+
\dfrac{\partial (P_yg(P))}{\partial P_y}\Bigg]\dfrac{1}{\omega-kv_TP_x}.
\eqno{(1.5)}
$$

Distribution function in the second approximation across the field
is  con\-stru\-cted
$$
f=f^{(2)}=f^{(0)}+f_1+f_2,
\eqno{(1.6)}
$$
where $f_1, f_2$ are given by equalities (1.4) and (1.6).

Let us find electric current density
$$
\mathbf{j}=e\int \mathbf{v}f \dfrac{2d^3p}{(2\pi\hbar)^3}.
\eqno{(1.7)}
$$

From equalities (1.4) -- (1.6) it is visible, that the vector
of density of a current has two nonzero components
$$
\mathbf{j}=(j_x,j_y,0).
$$
Here $j_y$ is the density of transversal current,
$$
j_y=e\int v_yf \dfrac{2d^3p}{(2\pi\hbar)^3}=
e\int v_yf_1 \dfrac{2d^3p}{(2\pi\hbar)^3}.
\eqno{(1.8)}
$$

This current is directed along an electromagnetic field, its density
it is defined only by the first approximation of function
of distribution. The second approximation of function of
distribution the contribution to current density
does not bring.

The density of the transversal current is defined by equality

$$
j_y=\dfrac{ie^2k_T^3}{2\pi^3m}E_y(x,t)\int\dfrac{P_y^2g(P)d^3P}
{\omega-kv_TP_x}.
$$

For density of the longitudinal current according to its definition it is had
$$
j_x=e\int v_xf\dfrac{2d^3p}{(2\pi\hbar)^3}=
e\int v_xf_2\dfrac{2d^3p}{(2\pi\hbar)^3}=
\dfrac{2ev_Tp_T^3}{(2\pi\hbar)^3}\int P_xf_2d^3P.
$$

By means of (1.6) from here it is received, that
$$
j_x=e^3E_y^2\dfrac{2mv_T^2}{(2\pi\hbar)^3\omega}\int
\Bigg[\dfrac{\partial (P_yg(P))}{\partial P_y}
+kv_TP_y^2\dfrac{\partial}{\partial P_x}
\Big(\dfrac{g(P)}{\omega-kv_TP_x}\Big)\Bigg]\dfrac{P_xd^3P}{\omega-kv_TP_x}.
\eqno{(1.9)}
$$

In the first integral from (1.9) internal integral on $P_y $ it is equal
to zero. In the second integral from (1.9) internal integral on $P_x $
it is calculated in parts
$$
\int\limits_{-\infty}^{\infty}\dfrac{\partial}{\partial P_x}
\Big(\dfrac{g(P)}{\omega-kv_TP_x}\Big)
\dfrac{P_xdP_x}{\omega-kv_TP_x}=
-\omega \int\limits_{-\infty}^{\infty}
\dfrac{g(P)dP_x}{(\omega-kv_TP_x)^3}.
$$

Hence, equality (1.9) becomes simpler
$$
j_x=-e^3E_y^2\dfrac{2mv_T^3k}{(2\pi\hbar)^3}\int \dfrac{g(P)P_y^2d^3P}
{(\omega-kv_TP_x)^3}=
\dfrac{e^3E_y^2q}{4\pi^3\hbar mv_T^2}\int\dfrac{g(P)P_y^2d^3P}
{(q P_x-\Omega)^3}.
\eqno{(1.10)}
$$
Here
$$
\Omega=\dfrac{\omega}{k_Tv_T},\qquad q=\dfrac{k}{k_T}.
$$

Equality (1.10) is reduced to one-dimensional integral
$$
j_x=\dfrac{e^3E_y^2q}{8\pi^2\hbar mv_T^2}\int\limits_{-\infty}^{\infty}
\dfrac{\ln(1+e^{\alpha-P_x^2})dP_x}{(q P_x-\Omega)^3}.
\eqno{(1.11)}
$$

Let us find numerical density (concentration) of particles of the plasma,
corresponding to  Fermi---Dirac  distribution
$$
N=\int f_0(P)\dfrac{2d^3p}{(2\pi\hbar)^3}=
\dfrac{8\pi p_T^3}{(2\pi\hbar)^3}\int\limits_{0}^{\infty}
\dfrac{e^{\alpha-P^2}P^2dP}{1+e^{\alpha-P^2}}=
\dfrac{k_T^3}{2\pi^2}l_0(\alpha),
$$
where $k_T$ is the thermal wave number, $k_T=\dfrac{mv_T}{\hbar}$,
$$
l_0(\alpha)=\int\limits_{0}^{\infty}\ln(1+e^{\alpha-P^2})dP.
$$

In expression before integral from (1.11) we will allocate the plasma
(Lang\-muir) frequency
$$
\omega_p =\sqrt {\dfrac{4\pi e^2N}{m}}
$$
and numerical density (concentration) $N $,
and last we will express through thermal wave number. We will receive
$$
{j_x}^{\rm long}={E_y^2}\dfrac{e\Omega_p^2}{p_T}\dfrac{q}{16\pi l_0(\alpha)}
\int\limits_{-\infty}^{\infty}
\dfrac{\ln(1+e^{\alpha-\tau^2})d\tau}{(q\tau-\Omega)^3},
$$
where
$\Omega_p=\dfrac{\omega_p}{k_Tv_T}=\dfrac{\hbar\omega_p}{mv_T^2}$
is the dimensionless plasma frequency.

This equality we will copy in the form
$$
j_x^{\rm long}=J_{\rm classic}(\Omega,q)\sigma_{l,tr}kE_y^2,
\eqno{(1.12)}
$$
where $\sigma_{l,tr}$ is the longitudinal--transversal conductivity,
$J_{\rm classic}(\Omega,q)$ is the dimensionless part of current,
$$
\sigma_{l,tr}=\dfrac{e\hbar}{p_T^2}\Big(\dfrac{\hbar \omega_p}{mv_T^2}\Big)^2=
\dfrac{e}{p_Tk_T}\Omega_p^2,
$$
$$
J_{\rm classic}(\Omega,q)=\dfrac{1}{16\pi l_0(\alpha)}
\int\limits_{-\infty}^{\infty}
\dfrac{\ln(1+e^{\alpha-\tau^2})d\tau}{(q\tau-\Omega)^3}.
$$

The integral from  dimensionless part of  current is calculated according to
to known Landau rule $$
\int\limits_{-\infty}^{\infty}
\dfrac{\ln(1+e^{\alpha-\tau^2})d\tau}{(q\tau-\Omega)^3}=
-i\dfrac{\pi}{2q^3}\Big[\ln(1+e^{\alpha-\tau^2})\Big]''
\Bigg|_{\tau=\Omega/q}+$$$$+{\rm V.p.}\int\limits_{-\infty}^{\infty}
\dfrac{\ln(1+e^{\alpha-\tau^2})d\tau}{(q\tau-\Omega)^3}.
$$\medskip

Symbol $ {\rm V.p.} $ before integral means, that integral
it is understood in sense of a principal value.

Let us introduce the transversal electromagnetic field
$$
\mathbf{E}_{\rm tr}=\mathbf{E}-\dfrac{\mathbf{k(Ek)}}{k^2}=
\mathbf{E}-\dfrac{\mathbf{q(Eq)}}{q^2}.
$$
Equality (1.12) can be written down in the invariant form
$$
\mathbf{j}^{\rm long}=J_{\rm classic}(\Omega,q)\sigma_{l,tr}{\bf k}{\bf E}_{tr}^2.
$$

Let us pass to consideration of the case of small values
of wave number. From expression
(1.10) at small values of wave number it is received
$$
j_x^{\rm classic}=-\dfrac{2e^3E_y^2mv_T^3k}{(2\pi\hbar)^3\omega^3}
\int g(P)P_y^2d^3P=-\dfrac{e^3E_y^2k_T^3l_0(\alpha)}{4\pi^2\omega^3}k=
$$
$$
=-\dfrac{1}{8\pi}\cdot \dfrac{e}{m\omega}
\Big(\dfrac{\omega_p}{\omega}\Big)^2kE_y^2.
\eqno{(1.13)}
$$

\section{Kinetic equation for Wigner fuction}

Shr\"{o}dinger equation
$$
i\hbar \dfrac{\partial \rho}{\partial t}=H\rho-{H^*}'\rho
$$
for  density matrix $ \rho $
under condition of calibration $ \rm div {\bf A} =0$ it will be transformed in
the kinetic equation \cite{Silin}

$$
\dfrac{\partial f}{\partial t}+
\mathbf{v}\dfrac{\partial f}{\partial {\bf r}}+W[f]=0,
\eqno{(2.1)}
$$
written down concerning quantum distribution Wigner function
$$
f(\mathbf{r},\mathbf{p},t)=\int
\rho(\mathbf{r}+\dfrac{\mathbf{a}}{2},\mathbf{r}-
\dfrac{\mathbf{a}}{2},t)e^{-i\mathbf{p}\mathbf{a}/\hbar}d^3a,
$$
besides
$$
\rho(\mathbf{R},\mathbf{R}',t)=\dfrac{1}{(2\pi \hbar)^3}
\int f(\dfrac{\mathbf{R}+\mathbf{R}'}{2}, \mathbf{p},t)
e^{i\mathbf{p}(\mathbf{R}-\mathbf{R}')/\hbar}d^3p.
$$

Here $H$ is the Hamilton operator, $H^*$ is the complex  conjugated
to $H$ operator, ${H^*}'$ is the complex  conjugated to
$H$ operator, acting on the shaded spatial variables
$\mathbf{r}'$.
The scalar potential is considered equal to zero.
Integral of Wigner is equal (see, example, \cite{Lat1}):
$$
W[f]=\iint\left\{-\dfrac{e}{2mc}
\Big[\mathbf{A}(\mathbf{r}+\dfrac{\mathbf{a}}{2},t)+
\mathbf{A}(\mathbf{r}-\dfrac{\mathbf{a}}{2},t)-
2\mathbf{A}(\mathbf{r},t)\Big]\dfrac{\partial f}{\partial {\bf r}}\right.-
$$\smallskip
$$
-\dfrac{ie}{ mc\hbar}\Big[
\mathbf{A}(\mathbf{r}+\dfrac{\mathbf{a}}{2},t)
-\mathbf{A}(\mathbf{r}-\dfrac{\mathbf{a}}{2},t)\Big]\mathbf{p'}
f+
$$\smallskip
$$+
\dfrac{i e^2}{2 mc^2\hbar}\Big[\mathbf{A}^2(\mathbf{r}+
\dfrac{\mathbf{a}}{2},t)-
\mathbf{A}^2(\mathbf{r}-\dfrac{\mathbf{a}}{2},t)\Big]f
\left. \right\}
e^{i(\mathbf{p'}-\mathbf{p})\mathbf{a}/\hbar}
\dfrac{d^3a\,d^3p'}{(2\pi\hbar)^3}.
$$

Vector potential of an electromagnetic field we take orthogonal
to direction of the wave vector $ {\bf k} $ ($ {\bf k A} =0$) in
the form of running harmonious wave
$$
{\bf A}({\bf r},t)={\bf A}_0e^{i({\bf kr}-\omega t)}.
$$

We transform the previous expression of Wigner integral
(see, example, \cite{Lat1}). We find that
$$
W[f]=-\mathbf{A}(\mathbf{r},t)\dfrac{e}{2mc}
\Big[\nabla %\dfrac{\partial}{\partial {\bf r}}
f(\mathbf{r},\mathbf{p}-\dfrac{\hbar \mathbf{k}}{2},t)+
\nabla f(\mathbf{r},\mathbf{p}+\dfrac{\hbar \mathbf{k}}{2},t)-
2\nabla f(\mathbf{r},\mathbf{p},t)\Big]-
$$\vspace{0.3cm}
$$
-\mathbf{A}(\mathbf{r},t)\dfrac{ie}{mc\hbar}
\Big\{\mathbf{p}\Big[f(\mathbf{r},
\mathbf{p}-\dfrac{\hbar \mathbf{k}}{2},t)-f(\mathbf{r},
\mathbf{p}+\dfrac{\hbar \mathbf{k}}{2},t)\Big]+
$$
$$
+\mathbf{A}^2(\mathbf{r},t)\dfrac{i e^2}{2 mc^2\hbar}
\Big[f(\mathbf{r},\mathbf{p}-{\hbar \mathbf{k}},t)-
f(\mathbf{r},\mathbf{p}+{\hbar \mathbf{k}},t)\Big].
\eqno{(2.2)}
$$

Let us enter local and absolute distributions of Fermi---Dirac
$$
f^{(0)}=f_0(\mathbf{r},{\bf C},t)=[1+\exp(C^2-\alpha)]^{-1},
$$
and
$$
f^{(0)}=f_0(P)=[1+\exp(P^2-\alpha)]^{-1}.
$$

Here
$$
{\bf C}\equiv\mathbf{C}(\mathbf{r},{\bf P},t)=\dfrac{\mathbf{v}}{v_T}=
{\bf P}-\dfrac{e}{cp_T}\mathbf{A}(\mathbf{r},t),
\qquad\qquad \alpha=\dfrac{\mu}{k_BT},
$$
$\mathbf{C}$ is the dimensionless electrons velocity,
$v_T={1}/{\sqrt{\beta}}$ is the thermal electrons velocity,
$\beta={m}/{2k_BT}$, ${\bf P}={{\bf p}}/{p_T}$ is the dimensionless
electrons momentum, $m$ is the electron mass, $k_B$ is the Boltzmann
constant, $T$ is the plasma temperature, $\mu$ is the chemical
potential of electronical gas, $\alpha$ is the dimen\-sion\-less
chemical potential.

Let us show, that the first composed in Wigner integral (2.2)
equals to  zero. We will notice, that according to problem statement
gradient of quantum  distribution function
it is proportional to the vector $ \mathbf {k} $:
$
\nabla f\sim \mathbf{k}.
$
Therefore
$$
\mathbf{A}(\mathbf{r},t)
\Big[\nabla f(\mathbf{r},\mathbf{p}-\dfrac{\hbar \mathbf{k}}{2},t)+
\nabla f(\mathbf{r},\mathbf{p}+\dfrac{\hbar \mathbf{k}}{2},t)-
2\nabla f(\mathbf{r},\mathbf{p},t)\Big]\sim \mathbf{Ak}=0.
$$\medskip

Thus, Wigner integral is equal
$$
W[f]=\mathbf{p A}\dfrac{ie}{mc\hbar}
\Big[f(\mathbf{r},\mathbf{p}+\dfrac{\hbar \mathbf{k}}{2},t)-
f(\mathbf{r},\mathbf{p}-\dfrac{\hbar \mathbf{k}}{2},t)\Big]+
$$
$$
-\mathbf{A}^2\dfrac{i e^2}{2 mc^2\hbar}
\Big[f(\mathbf{r},\mathbf{p}+\hbar \mathbf{k},t)-
f(\mathbf{r},\mathbf{p}-\hbar \mathbf{k},t)\Big].
$$\medskip

Let us return to the kinetic equation (2.1). We will consider
convectional derivative from this equation
$$
\mathbf{v}\dfrac{\partial f}{\partial \mathbf{r}}=
\Big(\dfrac{\mathbf{p}}{m}-\dfrac{e}{mc}\mathbf{A}(\mathbf{r},t)\Big)
\dfrac{\partial f}{\partial \mathbf{r}}.
$$

Thanks to conditions
$$
\dfrac{\partial f}{\partial \mathbf{r}}\sim \mathbf{k}, \;\qquad
\mathbf{A k}=0
$$
we receive that
$$
\mathbf{v}\dfrac{\partial f}{\partial \mathbf{r}}=\dfrac{\mathbf{p}}{m}
\dfrac{\partial f}{\partial \mathbf{r}}=v_T \mathbf{P}
\dfrac{\partial f}{\partial \mathbf{r}}.
$$

Let us solve further kinetic Wigner equation for quantum
distributions function
$$
\dfrac{\partial f}{\partial t}+v_T\mathbf{P}\dfrac{\partial f}{\partial
\mathbf{r}}+\dfrac{ie v_T\mathbf{PA}}{c\hbar}
\Big[f(\mathbf{r},\mathbf{P}+\dfrac{\mathbf{q}}{2},t)-
f(\mathbf{r},\mathbf{P}-\dfrac{\mathbf{q}}{2},t)\Big]-
$$
$$
-\mathbf{A}^2\dfrac{i e^2}{2 mc^2\hbar}
\Big[f(\mathbf{r},\mathbf{P}+\mathbf{q},t)-
f(\mathbf{r},\mathbf{P}-\hbar \mathbf{q},t)\Big]=0.
\eqno{(2.3)}
$$

Here
$$
\mathbf{q}=\dfrac{\hbar
\mathbf{k}}{p_T}=\dfrac{\mathbf{k}}{k_T},\;\qquad
k_T=\dfrac{p_T}{\hbar},
$$
$k_T$ is the thermal wave number, $\mathbf{q}$ is the dimensionless
wave number.

\section{Solution of Wigner equation}

Let us consider as small parametre size of vector potential
of electro\-mag\-ne\-tic field $ \mathbf {A} (\mathbf {r}, t) $. The solution
of the equations (2.3) we will be
to search by the method of consecutive approximations.

As the first approach for the solution we search in the form, linear
on vector potential concerning to absolute distribution of
Fermi---Dirac:
$$
f^{(1)}=f_0(P)+f_1,\qquad f_1\sim \mathbf{A}(\mathbf{r},t).
\eqno{(3.1)}
$$

As the first approximation we take linear concerning to vector
potential a part of Wigner equation
$$
\dfrac{\partial f}{\partial t}+v_T\mathbf{P}\dfrac{\partial f}{\partial
\mathbf{r}}+\dfrac{ie v_T\mathbf{P A}}{mc\hbar}
\Big[f_0(\mathbf{P}+\dfrac{\mathbf{q}}{2})-
f_0(\mathbf{P}-\dfrac{\mathbf{q}}{2})\Big]=0.
\eqno{(3.2)}
$$ \medskip
Here
$$
f_0\big(\mathbf{P}\pm\dfrac{\mathbf{q}}{2}\big)=
\Big[1+\exp\Big[\Big(\mathbf{P}\pm\dfrac{\mathbf{q}}{2}\Big)^2-\alpha\Big]
\Big]^{-1}.
$$

Substituting (3.1) in (3.2), we receive the equation
$$
-i(\omega-v_T\mathbf{ k P})f_1=-\dfrac{iev_T}{c\hbar}\mathbf{PA}
\Big[f_0(\mathbf{P}+\dfrac{\mathbf{q}}{2})-f_0(\mathbf{P}-
\dfrac{\mathbf{q}}{2})\Big].
$$

From this equation we receive
$$
f_1=\dfrac{ev_T}{c\hbar}\mathbf{PA}\dfrac{f_0(\mathbf{P}+
\mathbf{q}/{2})-f_0(\mathbf{P}-\mathbf{q}/{2})}
{\omega-v_T\mathbf{kP}}.
\eqno{(3.3)}
$$

Hence, as a first approximation according to (3.1) and (3.3)
the solution it is constructed
$$
f^{(1)}=f_0(P)+\dfrac{ev_T}{c\hbar}\mathbf{PA}\dfrac{f_0(\mathbf{P}+
\mathbf{q}/{2})-f_0(\mathbf{P}-\mathbf{q}/{2})}
{\omega-v_T\mathbf{kP}}.
\eqno{(3.4)}
$$

In the second approach we search for the solution in the form
$$
f^{(2)}=f^{(1)}+f_2, \qquad f_2 \sim \mathbf{A}^2(\mathbf{r},t).
$$

In the equation (2.3) in the first square bracket function $f $ we will replace
on $f^{(1)} $, and in the second square bracket function $f $ we will replace
on $f_0$, i.e. $f^{(2)} $ we search from the equation
$$
\dfrac{\partial f^{(2)}}{\partial t}+
v_T\mathbf{P}\dfrac{\partial f^{(2)}}{\partial
\mathbf{r}}+\dfrac{iev_T \mathbf{P A}}{c\hbar}
\Big[f^{(1)}(\mathbf{r},\mathbf{P}+\dfrac{\mathbf{q}}{2},t)-
f^{(1)}(\mathbf{r},\mathbf{P}-\dfrac{\mathbf{q}}{2},t)\Big]-
$$
$$
-\mathbf{A}^2\dfrac{i e^2}{2 mc^2\hbar}
\Big[f_0(\mathbf{P}+\mathbf{q})-f_0(\mathbf{P}-\mathbf{q})\Big]=0.
\eqno{(3.5)}
$$ \medskip

In this equation according to (3.4)
$$
f^{(1)}\Big(\mathbf{P}+\dfrac{\mathbf{q}}{2}\Big)=
f_0\Big(\mathbf{P}+\dfrac{\mathbf{q}}{2}\Big)
+\dfrac{ev_T}{c\hbar}\Big(\mathbf{P}+\dfrac{\mathbf{q}}{2}\Big)\mathbf{A}
\dfrac{f_0(\mathbf{P+q})-f_0(P)}{\omega-v_T\mathbf{k(P+q}/2)},
$$
$$
f^{(1)}\Big(\mathbf{P}-\dfrac{\mathbf{q}}{2}\Big)=
f_0\Big(\mathbf{P}-\dfrac{\mathbf{q}}{2}\Big)
+\dfrac{ev_T}{c\hbar}\Big(\mathbf{P}-\dfrac{\mathbf{q}}{2}\Big)\mathbf{A}
\dfrac{f_0(P)-f_0(\mathbf{P-q})}{\omega-v_T\mathbf{k(P-q}/2)},
$$\medskip

We notice that
$$
\mathbf{A q}=0,\;\quad\text{because}\;\quad \mathbf{A q}\sim \mathbf{A k}=0.
$$

Therefore the previous two parities become simpler
$$
f^{(1)}\Big(\mathbf{P}-\dfrac{\mathbf{q}}{2}\Big)=
f_0\Big(\mathbf{P}-\dfrac{\mathbf{q}}{2}\Big)
+\dfrac{ev_T}{c\hbar}(\mathbf{PA})
\dfrac{f_0(P)-f_0(\mathbf{P-q})}{\omega-v_T\mathbf{k(P-q}/2)},
$$\medskip
$$
f^{(1)}\Big(\mathbf{P}+\dfrac{\mathbf{q}}{2}\Big)=
f_0\Big(\mathbf{P}+\dfrac{\mathbf{q}}{2}\Big)
+\dfrac{ev_T}{c\hbar}\big(\mathbf{PA}\big)
\dfrac{f_0(\mathbf{P+q})-f_0(P)}{\omega-v_T\mathbf{k(P+q}/2)}.
$$\medskip

The equation (3.5) we will copy in an explicit form
$$
\dfrac{\partial f^{(1)}}{\partial t}+v_T\mathbf{P}\dfrac{\partial f^{(1)}}
{\partial \mathbf{r}}+\dfrac{iev_T\mathbf{PA}}{c\hbar}\Big[f_0(\mathbf{P}+
\dfrac{\mathbf{q}}{2})-f_0(\mathbf{P}-\dfrac{\mathbf{q}}{2})\Big]+
$$
$$
+\dfrac{\partial f_2}{\partial t}+v_T\mathbf{P}\dfrac{\partial f_2}
{\partial \mathbf{r}}+i\dfrac{e^2v_T^2(\mathbf{PA})^2}{c^2\hbar^2}
\Big[\dfrac{f_0(\mathbf{P+q})-f_0(P)}{\omega-v_T\mathbf{k(P+q}/2)}-
$$
$$
-\dfrac{f_0(P)-f_0(\mathbf{P-q})}{\omega-v_T\mathbf{k(P-q}/2)}\Big]-
\dfrac{ie^2\mathbf{A}^2}{2mc^2\hbar}\Big[f_0(\mathbf{P+q})-
f_0(\mathbf{P-q})\Big]=0.
$$

First three composed in this equation give zero agree
to the equation (3.2). The rest part of equation leads
to equality
$$
-2i(\omega-v_T\mathbf{kP})f_2=-i\dfrac{e^2v_T^2(\mathbf{PA})^2}{c^2\hbar^2}
\Big[\dfrac{f_0(\mathbf{P+q})-f_0(P)}{\omega-v_T\mathbf{k(P+q}/2)}-
$$
$$
-\dfrac{f_0(P)-f_0(\mathbf{P-q})}{\omega-v_T\mathbf{k(P-q}/2)}\Big]+
\dfrac{ie^2\mathbf{A}^2}{2mc^2\hbar}\Big[f_0(\mathbf{P+q})-
f_0(\mathbf{P-q})\Big].
$$

From here we find that
$$
f_2=\dfrac{e^2v_T^2(\mathbf{PA})^2}{2c^2\hbar^2(\omega-v_T\mathbf{kP})}
\Big[\dfrac{f_0(\mathbf{P+q})-f_0(P)}{\omega-v_T\mathbf{k(P+q}/2)}-
$$
$$
-\dfrac{f_0(P)-f_0(\mathbf{P-q})}{\omega-v_T\mathbf{k(P-q}/2)}\Big]-
\dfrac{e^2\mathbf{A}^2}{4mc^2\hbar}\dfrac{f_0(\mathbf{P+q})-
f_0(\mathbf{P-q})}{\omega-v_T\mathbf{kP}}.
\eqno{(3.6)}
$$

So, quantum  distribution function of Wigner is constructed and
it is defined by equalities (3.1), (3.3) and (3.6):
$f=f_0(P) +f_1+f_2$, or, is more detailed
$$
f=f_0(P)+\dfrac{ev_T}{c\hbar}\mathbf{PA}\dfrac{f_0(\mathbf{P}+
\mathbf{q}/{2})-f_0(\mathbf{P}-\mathbf{q}/{2})}
{\omega-v_T\mathbf{kP}}+
%$$
%$$
\dfrac{e^2v_T^2(\mathbf{PA})^2}{2c^2\hbar^2(\omega-v_T\mathbf{kP})}\times
$$$$\times\Big[\dfrac{f_0(\mathbf{P+q})-f_0(P)}{\omega-v_T\mathbf{k(P+q}/2)}+
\dfrac{f_0(\mathbf{P-q})-f_0(P)}{\omega-v_T\mathbf{k(P-q}/2)}\Big]-$$$$-
\dfrac{e^2\mathbf{A}^2}{4mc^2\hbar}\dfrac{f_0(\mathbf{P+q})-
f_0(\mathbf{P-q})}{\omega-v_T\mathbf{kP}}.
$$

\section{Electrical current in quantum plasma}

By definition, the electric current density is equal
$$
\mathbf{j}(\mathbf{r},t)=e\int \mathbf{v}(\mathbf{r},\mathbf{p},t)
f(\mathbf{r},\mathbf{p},t)
\dfrac{2\,d^3p}{(2\pi\hbar)^3}.
\eqno{(4.1)}
$$

Substituting in equality (4.1) obvious expression for velocity
$$
\mathbf{v}(\mathbf{r},\mathbf{P},t)=
v_T\mathbf{P}-\dfrac{e \mathbf{A}(\mathbf{r},t)}{mc},
$$
and distribution function according to equality $f=f_0(P)+f_1+f_2$.

Leaving linear and square-law expressions concerning
vector potential of a field, we receive
$$
\mathbf{j}=\dfrac{2ep_T^3}{(2\pi\hbar)^3}\int \Big[v_T\mathbf{P}f_1-
\dfrac{e}{mc}\mathbf{A}f_0(P)\Big]d^3P+
$$
$$
+\dfrac{2ep_T^3}{(2\pi\hbar)^3}\int \Big[v_T\mathbf{P}f_2-
\dfrac{e}{mc}\mathbf{A}f_1\Big]d^3P.
\eqno{(4.2)}
$$ \medskip

Let us show, that the formula (4.2) for electric current density
contains two nonzero components: $ \mathbf {j} = (j_x, j_y, 0) $.
One component $j_y $ it is linear on potential of an electromagnetic field and
is directed lengthways field. It is the known formula for
electric current density, so-called "transversal current".
The second  component $j_x $ is quadratic on potential of
field also and it is directed along the wave
vector. It is "longitudinal current".

The first composed in (4.2) is linear on vector potential
expression, and second is square-law. We will write out these composed in
obvious form
$$
\mathbf{j}^{\rm linear}=\dfrac{2ep_T^3}{(2\pi\hbar)^3}\int \Bigg[
\dfrac{ev_T^2}{c\hbar}\mathbf{P(PA)}\dfrac{f_0(\mathbf{P+q}/2)-
f_0(\mathbf{P-q}/2)}{\omega-v_T\mathbf{kP}}-\dfrac{e\mathbf{A}}{mc}f_0(P)
\Bigg]d^3P
\eqno{(4.3)}
$$
and
$$
\mathbf{j}^{\rm quadr}=\dfrac{2ep_T^3}{(2\pi\hbar)^3}\int \Bigg[
-\dfrac{e^2v_T\mathbf{A(PA)}}{mc^2\hbar}
\big[f_0(\mathbf{P}+\dfrac{{\bf q}}{2})-
f_0(\mathbf{P}-\dfrac{{\bf q}}{2})\big]+
$$
$$
+\dfrac{e^2v_T^3\mathbf{P(PA)}^2}{2c^2\hbar^2}
\Big[\dfrac{f_0(\mathbf{P+q})-f_0(P)}{\omega-v_T\mathbf{k(P+q}/2)}-
\dfrac{f_0(P)-f_0(\mathbf{P-q})}{\omega-v_T\mathbf{k(P-q}/2)}\Big]-
$$
$$
-\dfrac{e^2v_T\mathbf{PA}^2}{4mc^2\hbar}\big[f_0(\mathbf{P+q})-
f_0(\mathbf{P-q})\big]\Bigg]
\dfrac{d^3P}{\omega-v_T\mathbf{kP}}.
\eqno{(4.4)}
$$

Expression (4.3) is linear expression of
the electric current, found, in particular, in our previous
work \cite{Lat1}. This vector expression contains only one
component, directed along the electromagnetic
fields. Really, if a wave vector to direct along an axis
$x $ i.e. to take $ \mathbf {k} =k (1,0,0) $, and potential electromagnetic
fields to direct along an axis $y $, i.e. to take
$$
\mathbf {A} (\mathbf {r}, t) = (0, A_y (x, t), 0),
$$
from the formula (4.3) we receive
$$
j_y^{\rm linear}=-\dfrac{2e^2p_T^3A_y}{(2\pi\hbar)^3mcq}
\int\Big(\dfrac{f_0(P_x+q/2)-f_0(P_x-q/2)}
{P_x-\Omega/q}P_y^2+qf_0(P)\Big)d^3P.
\eqno{(4.5)}
$$
Here
$$
f_0(P_x\pm q/2)=\Big[1+e^{(P_x\pm q/2)^2+P_y^2+P_z^2-\alpha}\Big]^{-1}, \qquad
\Omega=\dfrac{\omega}{k_Tv_T}.
$$

Let us consider expression for an electric current (4.4),
proportional to a square of potential of an electromagnetic field.
Let us notice, that the first composed in this expression is equal to zero.
Hence, this expression becomes simpler
$$
\mathbf{j}^{\rm quadr}=\dfrac{2ep_T^3}{(2\pi\hbar)^3}\int \Bigg[
\dfrac{e^2v_T^3\mathbf{P(PA)}^2}{2c^2\hbar^2(\omega-v_T\mathbf{kP})}
\Big[\dfrac{f_0(\mathbf{P+q})-f_0(P)}{\omega-v_T\mathbf{k(P+q}/2)}+$$$$+
\dfrac{f_0(\mathbf{P-q})-f_0(P)}{\omega-v_T\mathbf{k(P-q}/2)}\Big]-
\dfrac{e^2v_T\mathbf{PA}^2}{4mc^2\hbar}\dfrac{f_0(\mathbf{P+q})-
f_0(\mathbf{P-q})}{\omega-v_T\mathbf{kP}}\Bigg]d^3P.
\eqno{(4.6)}
$$

Let us notice, that vector expression (4.6) contains one nonzero
of the electric current component, directed along the wave
vector
$$
{j_x}^{\rm quadr}=
\dfrac{e^3p_T^3A_y^2}{(2\pi\hbar)^3c^2m^2v_T}\int \Bigg[
\Big[\dfrac{f_0(P_x+q)-f_0(P)}{qP_x+q^2/2-\Omega}+
\dfrac{f_0(P_x-q)-f_0(P)}{qP_x-q^2/2-\Omega}\Big]P_y^2
+
$$
$$
+\dfrac{1}{2}\big[f_0(P_x+q)-f_0(P_x-q)\big]
\Bigg]\dfrac{P_xd^3P}{qP_x-\Omega}.
\eqno{(4.7)}
$$

Let us lead to a kind convenient for calculations, the formula (4.7) for
density of a longitudinal current.

Let's consider the first integral from (4.7). We will calculate the internal
integrals in  plane $ (P_y, P_z) $, passing to polar coordinates
$$
\int f_0(P_x\pm q)P_y^2dP_ydP_z=
\pi \int\limits_{0}^{\infty}
\rho\ln(1+e^{\alpha-(P_x\pm q)^2-\rho^2})d\rho,
$$
$$
\int f_0(P_x\pm q)dP_ydP_z=\pi\ln(1+e^{\alpha-(P_x\pm q)^2}).
$$

Thus, size of density of generated longitudinal current into
quantum plasma it is equal
$$
j_{x}^{\rm quant}=
\dfrac{\pi e^3p_T^3A_y^2}{(2\pi\hbar)^3m^2c^2v_T}
\int\limits_{-\infty}^{\infty}\Bigg[
\dfrac{L(P_x+q,P_x)}{qP_x+q^2/2-\Omega}+\dfrac{L(P_x-q,P_x)}{qP_x-q^2/2-\Omega}+
$$
$$
+\dfrac{1}{2}\ln\dfrac{1+e^{\alpha-(P_x+q)^2}}
{1+e^{\alpha-(P_x-q)^2}}\Bigg]\dfrac{P_x dP_x}{qP_x-\Omega}.
\eqno{(4.8)}
$$

Here
$$
L(P_x\pm q,P_x)=\int\limits_{0}^{\infty}
\rho\ln\dfrac{1+e^{\alpha-(P_x\pm q)^2-\rho^2}}{1+e^{\alpha-P_x^2-\rho^2}}d\rho.
$$

Let us transform the expression standing in integral (4.8). At first
let us pass from potential to intensity of a field
$A_y = - (ic/\omega) E_y $. We will receive
$$
\dfrac{\pi e^3p_T^3A_y^2}{(2\pi\hbar)^3m^2c^2v_T}
=-\dfrac{\pi e^3k_T^3E_y^2}{8\pi^3m^2v_T\omega^2}.
$$

Let us transform this expression by means of expression for the thermal
wave numbers $k_T $. We receive, that
$$
-\dfrac{\pi e^3k_T^3E_y^2}{8\pi^3m^2v_T\omega^2}=
-\dfrac{e^3NE_y^2}{4l_0(\alpha)m^2v_T\omega^2}=
-\dfrac{1}{16\pi}\cdot\dfrac{e}{mv_T}
\Big(\dfrac{\omega_p}{\omega}\Big)^2E_y^2=
$$
$$
=-\dfrac{1}{16\pi}\dfrac{e}{p_T}\Big(\dfrac{\Omega_p}{\Omega}\Big)^2E_y^2
=-\dfrac{1}{16\pi l_0(\alpha)\Omega^2}\dfrac{e\Omega_p^2}{p_T}E_y^2
=-\dfrac{1}{16\pi l_0(\alpha)q \Omega^2}\sigma_{l,tr}kE_y^2,
$$
where the quantity of longitudinal--transversal coductivity $\sigma_{l,tr}$
was introduced in part 1: $\sigma_{l,tr}=e\Omega_p^2/(p_Tk_T)$.

Now equality (4.8) we will present in the form
$$
j_x^{\rm quant}=J_{\rm quant}(\Omega,q)\sigma_{l,tr}kE_y^2(x,t).
\eqno{(4.9)}
$$

In  (4.9) $J_{\rm quant}(\Omega,q)$ is the density of dimensionless
longitudinal current,
$$
J_{\rm quant}(\Omega,q)=-\dfrac{1}{16\pi l_0(\alpha)q\Omega^2}
\int\limits_{-\infty}^{\infty}\Bigg[
\dfrac{L(\tau+q,\tau)}{q\tau+q^2/2-\Omega}+
\dfrac{L(\tau-q,\tau)}{q\tau-q^2/2-\Omega}+
$$
$$
+\dfrac{1}{2}\ln\dfrac{1+e^{\alpha-(\tau+q)^2}}
{1+e^{\alpha-(\tau-q)^2}}\Bigg]\dfrac{\tau d\tau}{q\tau-\Omega}.
\eqno{(4.10)}
$$

Let us transform integral from (4.10) from the first composed
$$
J_1=\int\limits_{-\infty}^{\infty}\Bigg[
\dfrac{L(\tau+q,\tau)}{q\tau+q^2/2-\Omega}+
\dfrac{L(\tau-q,\tau)}{q\tau-q^2/2-\Omega}\Bigg]
\dfrac{\tau d\tau}{q\tau-\Omega}=J_2+J_3.
$$

Here
$$
J_2=\int\limits_{-\infty}^{\infty}
\dfrac{L(\tau+q,\tau)\tau d\tau}{(q\tau+q^2/2-\Omega)(q\tau-\Omega)},
$$
$$
J_3=\int\limits_{-\infty}^{\infty}
\dfrac{L(\tau-q,\tau)\tau d\tau}{(q\tau-q^2/2-\Omega)(q\tau-\Omega)}.
$$

In integral $J_2$ we will make variable replacement $ \tau\to\tau-q/2$, and
in integral $J_3$ we will make replacement $ \tau\to \tau+q/2$.
As result it is received
$$
J_2=\int\limits_{-\infty}^{\infty}
\dfrac{L(\tau+q/2,\tau-q/2)(\tau-q/2)d\tau}{(q\tau-\Omega)(q\tau-q^2/2-\Omega)},
$$
$$
J_3=\int\limits_{-\infty}^{\infty}
\dfrac{L(\tau-q/2,\tau+q/2)(\tau+q/2)}
{(q\tau-\Omega)(q\tau+q^2/2-\Omega)}.
$$

We notice that
$$
L(\tau-q/2,\tau+q/2)=\int\limits_{0}^{\infty}\rho
\ln\dfrac{1+e^{\alpha-(\tau-q/2)^2-\rho^2}}
{1+e^{\alpha-(\tau+q/2)^2-\rho^2}}d\rho=$$$$=-
\int\limits_{0}^{\infty}\rho
\ln\dfrac{1+e^{\alpha-(\tau+q/2)^2-\rho^2}}
{1+e^{\alpha-(\tau-q/2)^2-\rho^2}}d\rho=-L(\tau+q/2,\tau-q/2).
$$

Hence, integral $J_1$ equals
$$
J_1=\int\limits_{-\infty}^{\infty}\dfrac{L(\tau+q/2,\tau-q/2)}{q\tau-\Omega}
\Big[\dfrac{\tau-q/2}{q\tau-q^2/2-\Omega}-
\dfrac{\tau+q/2}{q\tau+q^2/2-\Omega}\Big]d\tau=
$$
$$
=q\Omega\int\limits_{-\infty}^{\infty}\dfrac{L(\tau+q/2,\tau-q/2)d\tau}
{(q\tau-\Omega)[(q\tau-\Omega)^2-q^4/4]}.
$$

Let us calculate the second integral from (4.10)
$$
J_4=\dfrac{1}{2}\int\limits_{-\infty}^{\infty}
\ln\dfrac{1+e^{\alpha-(\tau+q)^2}}
{1+e^{\alpha-(\tau-q)^2}}\dfrac{\tau d\tau}{q\tau-\Omega}=
$$
$$
=\dfrac{1}{2q}\int\limits_{-\infty}^{\infty}\ln\dfrac{1+e^{\alpha-(\tau+q)^2}}
{1+e^{\alpha-(\tau-q)^2}}\Big(1+\dfrac{\Omega}{q\tau-\Omega}\Big)d\tau=
$$
$$
=\dfrac{\Omega}{2q}\int\limits_{-\infty}^{\infty}\ln\dfrac{1+e^{\alpha-(\tau+q)^2}}
{1+e^{\alpha-(\tau-q)^2}}\dfrac{d\tau}{q\tau-\Omega}=
$$
$$
=\dfrac{\Omega}{2q}\int\limits_{-\infty}^{\infty}\ln(1+e^{\alpha-x^2})
\Big[\dfrac{1}{qx-q^2-\Omega}-\dfrac{1}{qx+q^2-\Omega}\Big]dx=
$$
$$
=\Omega q\int\limits_{-\infty}^{\infty}\dfrac{\ln(1+e^{\alpha-x^2})dx}
{(qx-\Omega)^2-q^4/4}dx.
$$

Finally, integral (4.10) equals
$$
J_{\rm quant}(\Omega,q)=-\dfrac{1}{16\pi l_0(\alpha)\Omega^2}\Bigg[
\int\limits_{-\infty}^{\infty}\dfrac{L(\tau+q/2,\tau-q/2)d\tau}
{(q\tau-\Omega)[(q\tau-\Omega)^2-q^4/4]}+$$$$+
\int\limits_{-\infty}^{\infty}\dfrac{\ln(1+e^{\alpha-x^2})dx}
{(qx-\Omega)^2-q^4/4}dx\Bigg].
\eqno{(4.11)}
$$

At calculation singular integral from (4.10), which not writing out
let us designate through $I(\Omega, q) $, it is necessary to take advantage
known Landau rule. Then
$$
I(\Omega,q)=\Re I(\Omega,q)+i\Im I(\Omega,q).
$$
Here
$$
\Re I(\Omega,q)={\rm V.p.}
\int\limits_{-\infty}^{\infty}\dfrac{L(\tau+q/2,\tau-q/2)d\tau}
{(q\tau-\Omega)[(q\tau-\Omega)^2-q^4/4]}+$$$$+{\rm V.p.}
\int\limits_{-\infty}^{\infty}\dfrac{\ln(1+e^{\alpha-x^2})dx}
{(qx-\Omega)^2-q^4/4}dx,
$$
Symbol ${\rm V.p.}$  means principal value of integral also,
$$
\Im I(\Omega,q)=-\dfrac{\pi}{q^4}
\Bigg\{(q^2-2\Omega)
L\Big(\dfrac{q}{2}+\dfrac{\Omega}{q},-\dfrac{q}{2}+\dfrac{\Omega}{q}\Big)+
$$$$+
(q^2+2\Omega)L\Big(-\dfrac{q}{2}+\dfrac{\Omega}{q},
\dfrac{q}{2}+\dfrac{\Omega}{q}\Big)+
$$
$$
+2\Omega\Big[L\Big(\dfrac{\Omega}{q}+q,\dfrac{\Omega}{q}\Big)-
L\Big(\dfrac{\Omega}{q}-q,\dfrac{\Omega}{q})\Big]+
\dfrac{\Omega q^2}{2}\ln\dfrac{1+e^{\alpha-(\Omega/q+q)^2}}
{1+e^{\alpha-(\Omega/q-q)^2}}\Bigg\}.
$$

Considering antisymmetry of function $L (X, Y) $ on the variables
($L (X, Y) =-L (Y, X) $), we will simplify expression for an imaginary part
of density of the dimensionless current. It is as a result received, that
$$
\Im I(\Omega,q)=-\dfrac{\pi \Omega}{q^4}
\Bigg[-4L\Big(\dfrac{q}{2}+\dfrac{\Omega}{q},
-\dfrac{q}{2}+\dfrac{\Omega}{q}\Big)+
$$
$$
+2\Omega\Big[L\Big(\dfrac{\Omega}{q}+q,\dfrac{\Omega}{q}\Big)-
L\Big(\dfrac{\Omega}{q}-q,\dfrac{\Omega}{q}\Big)\Big]+
\dfrac{\Omega q^2}{2}\ln\dfrac{1+e^{\alpha-(\Omega/q+q)^2}}
{1+e^{\alpha-(\Omega/q-q)^2}}\Bigg].
$$

Equality (4.10) for density of a longitudinal current we will present
into the vector form
$$
{\bf j}^{\rm quant}=J_{\rm quant}(\Omega,q)\sigma_{l,tr}k{\bf E}_{tr}^2.
$$

Let us show, that at small values of wave number density
longi\-tu\-di\-nal current both in quantum and in classical plasma
coincide.

According to (4.7) at small $q $ after
linearization $f_0 (P_x\pm q) $ it is received
$$
j_x^{\rm quant}=\dfrac{2e^3p_T^3A_y^2q}{(2\pi\hbar)^3c^2m^2v_T\Omega}
\int g(P)P_x^2d^3P=\dfrac{2\pi e^3p_T^3A_y^2l_0(\alpha)}
{(2\pi\hbar)^3c^2m^2\omega}k.
$$

Let us take advantage of relation  of thermal wave number with the nume\-rical
density, and also relation of potential and intensity of
electromagnetic field. We receive, that
$$
j_x^{\rm quant}=-\dfrac{1}{8\pi}\cdot \dfrac{e}{m\omega}\
\Big(\dfrac{\omega_p}{\omega}\Big)^2kE_y^2.
\eqno{(4.11)}
$$

Expression (4.11) in accuracy coincides with (1.12). These expressions
let us copy in the vector kind
$$
{\bf j}^{\rm long}=-\dfrac{1}{8\pi}\cdot\dfrac{e}{m\omega}
\Big(\dfrac{\omega_p}{\omega}\Big)^2{\bf k}{\bf E}_{tr}^2.
$$

\section{Conclutions}

Graphic investigations of sizes of density of  longitudinal current
we will spend for case:  chemical potential equals to zero ($ \alpha=0$).
Curves {\it 1} answer to classical plasma, curves {\it 2} to quantum.

On figs. 1 and 2 we will present behaviour real (fig. 1) and imaginary
(fig. 2) parts of density of dimensionless longitudinal current
at $ \Omega=0.5$
Depending on dimensionless wave number $q $. At decrease
parametre $ \Omega $ curves {\it 1} and {\it 2} approach and
become indiscernible.

On fig. 3 and 4 we will represent behaviour real (fig. 3) and imaginary
(fig. 4) density parts of
longitudinal current depending on dimensionless wave number $q $
in the case $ \Omega=1$. From these drawings it is visible, that with growth
dimensionless frequency of oscillations of an electromagnetic field of ordinate
graphics quickly decrease.

On fig. 5 and 6 we will represent behaviour real (fig. 5) and imaginary
(fig. 6) parts of
longitudinal current depending on dimensionless frequency of oscillations
of electromagnetic field $ \Omega $ in the case $q=0.3$. At reduction
dimensionless wave number $q $ curves {\it 1} and {\it 2}
approximate and at small $q $ practically coincide.

On figs. 7--10 we sonsider behaviour of classical plasmas.

On fig. 7 and 8 the behaviour real (fig. 7) and imaginary (fig. 8)
is represented  parts of
longitudinal current depending on dimensionless wave number,
$ \Omega $. Curves 1,2,3 accordingly answer values
dimensionless chemical potential $ \alpha =-5,0,3$.

On fig. 9 and 10 we will represent behaviour real (fig. 9) and
imaginary (fig. 10) parts of  longitudinal current depending on
dimensionless wave number $q $ in a case $ \Omega=1$. Curves
1,2,3 answer according to values of the dimensionless
chemical potential $ \alpha =-5,0,3$.

In the present work the account of nonlinear character of
interaction electromagnetic field with classical
and quantum plasma is considered.
It has appeared, that
the account of nonlinearity of an electromagnetic field finds out
generating of an electric current, orthogonal to a direction
of electro\-mag\-ne\-tic field.

Further authors purpose to consider new problems about oscillations
plas\-mas and about skin-effect with use square-law onto
potential of expansion of  distribution  function.

\clearpage

\begin{figure}[t]\center
\includegraphics[width=16.0cm, height=10cm]{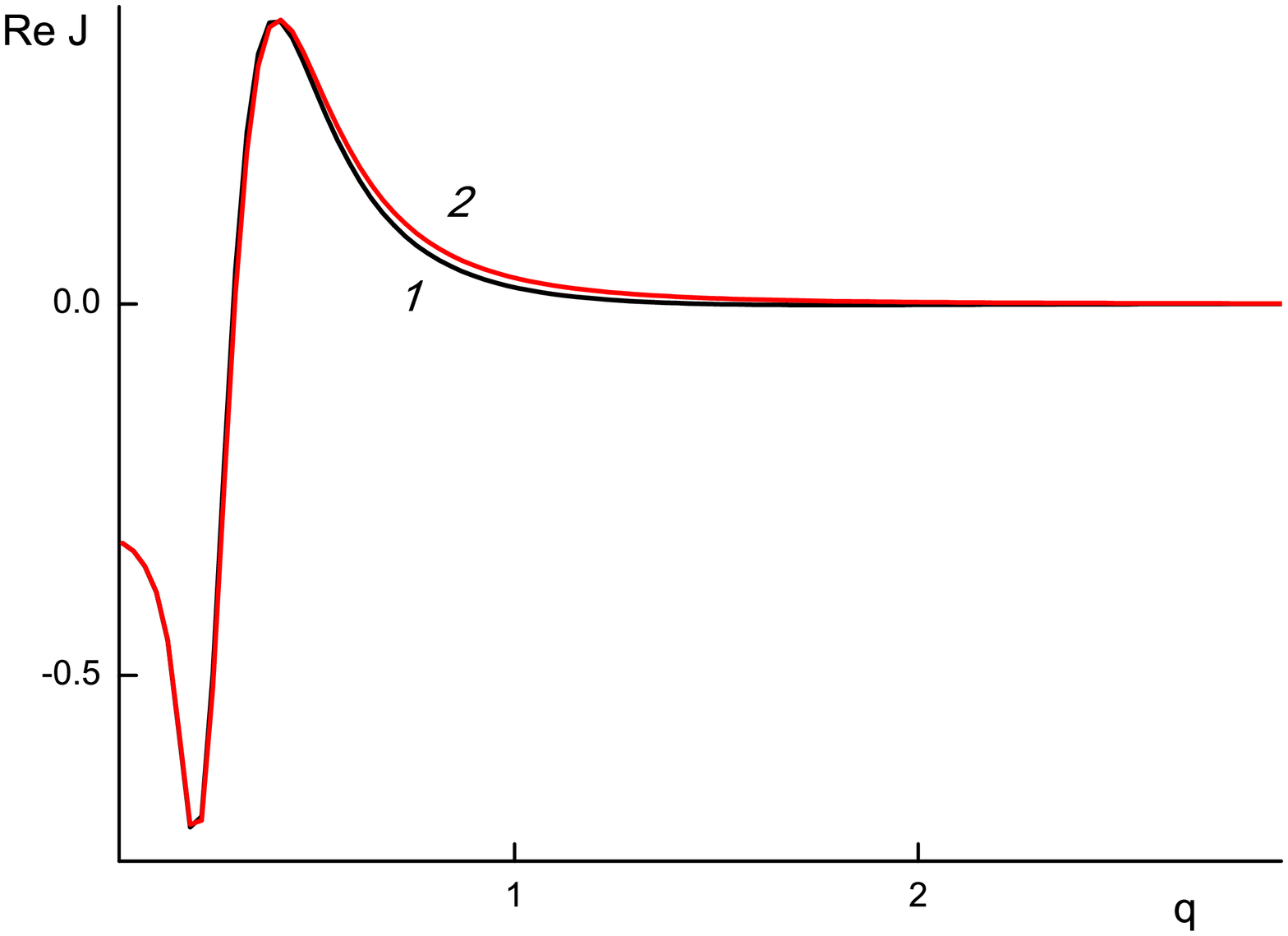}
{{\bf Fig. 1.} Real part of density of dimensionless current, $\Omega=0.5$.
}
\includegraphics[width=16.0cm, height=10cm]{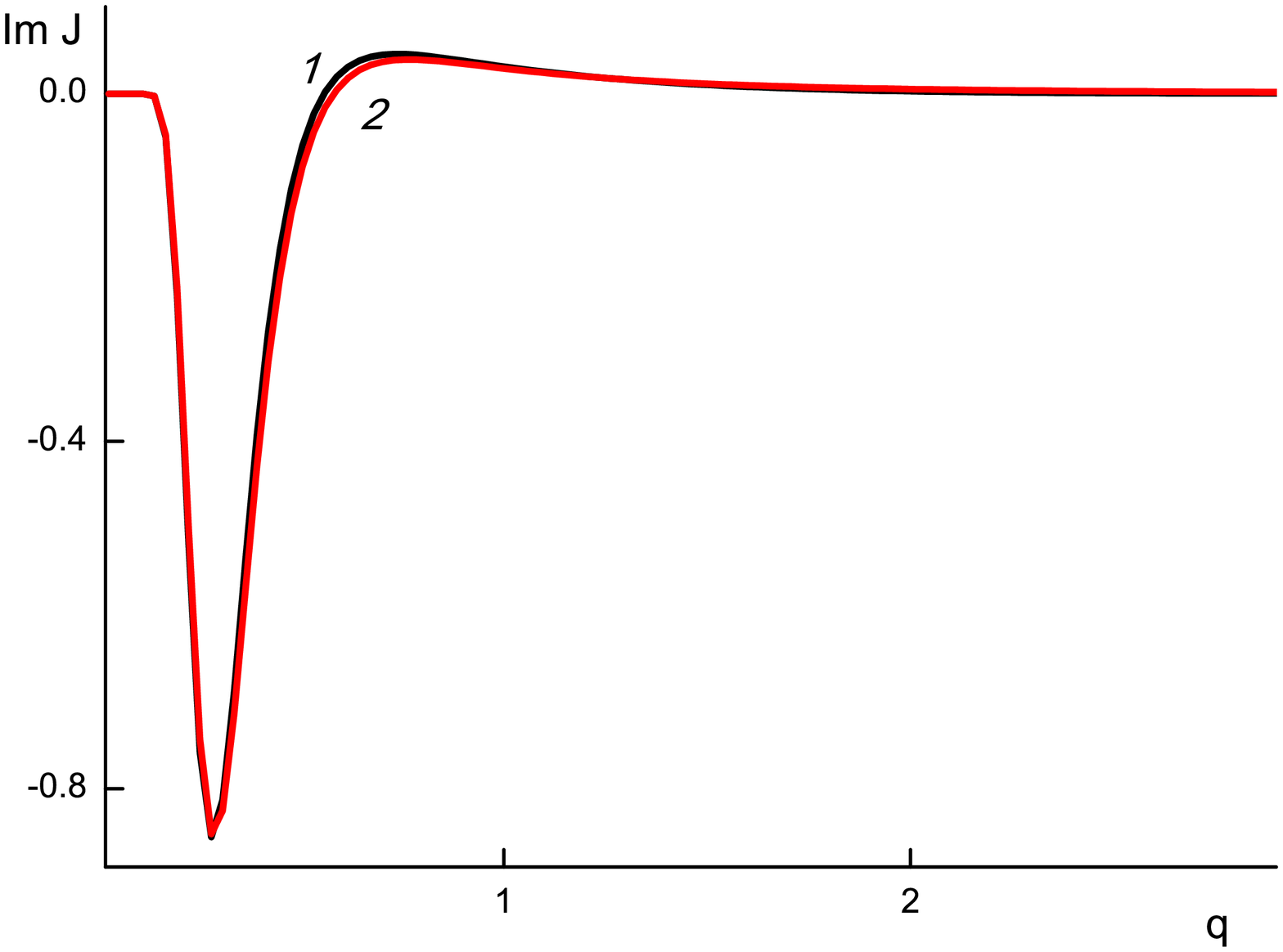}
{{\bf Fig. 2.} Imaginary part of density of dimensionless current,
$\Omega=0.5$.}
\end{figure}

\begin{figure}[th]\center
\includegraphics[width=16.0cm, height=10cm]{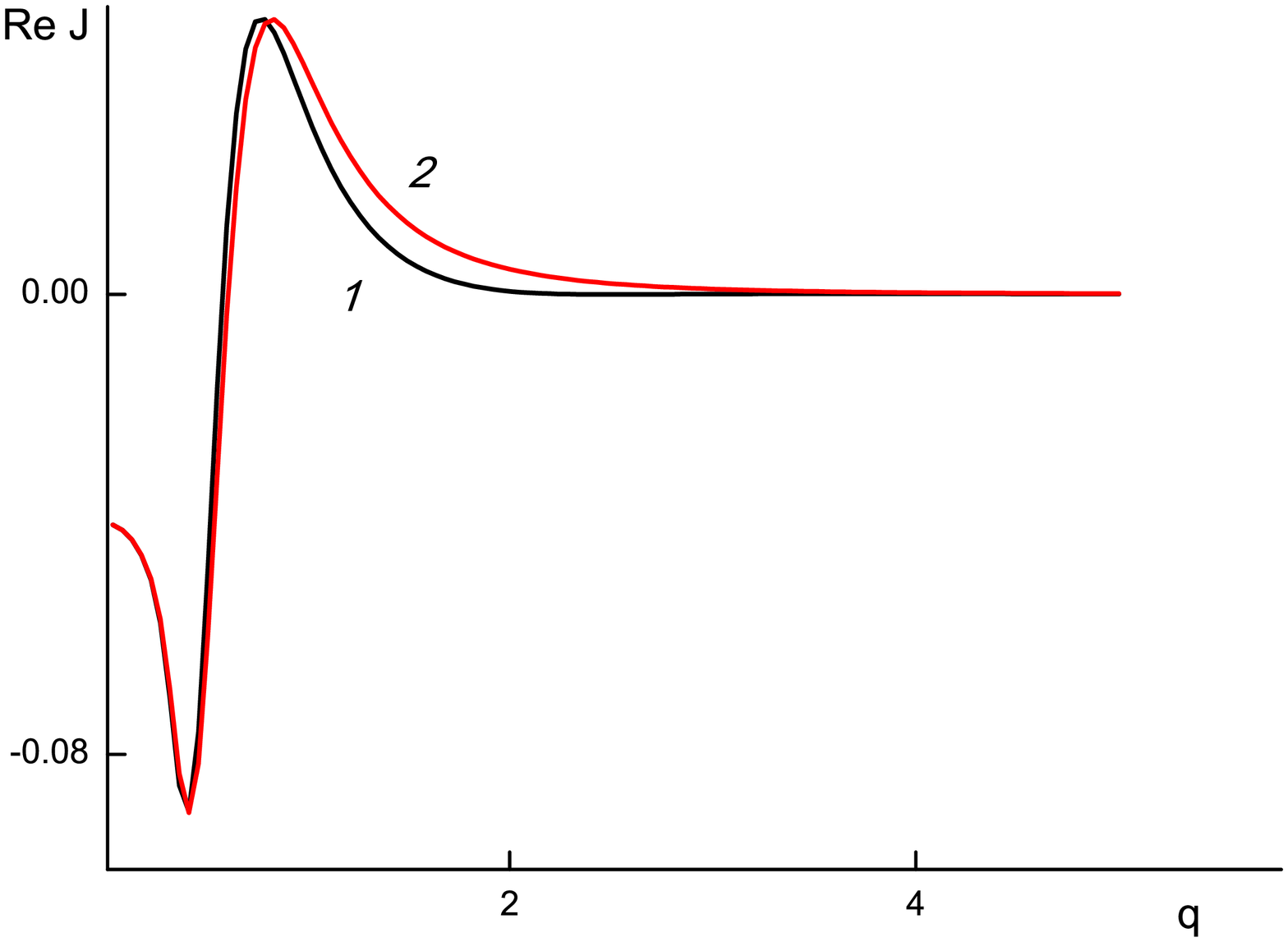}
{{\bf Fig. 3.} Real part of density of dimensionless current, $\Omega=1$.}
\includegraphics[width=16.0cm, height=10cm]{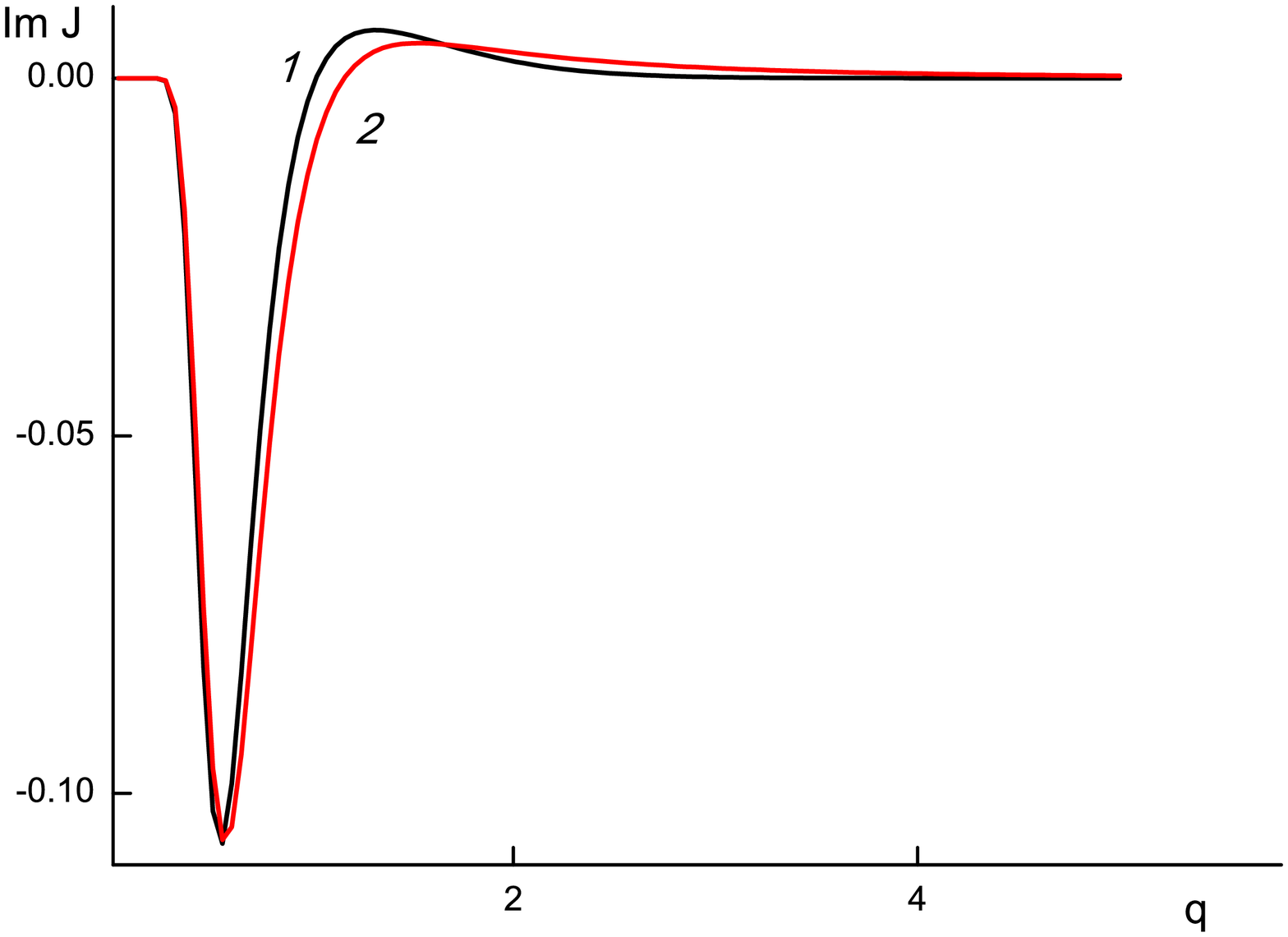}
{{\bf Fig. 4.} Imaginary part of density of dimensionless current, $\Omega=1$.
}
\end{figure}

\begin{figure}[thp]\center
\includegraphics[width=16.0cm, height=10cm]{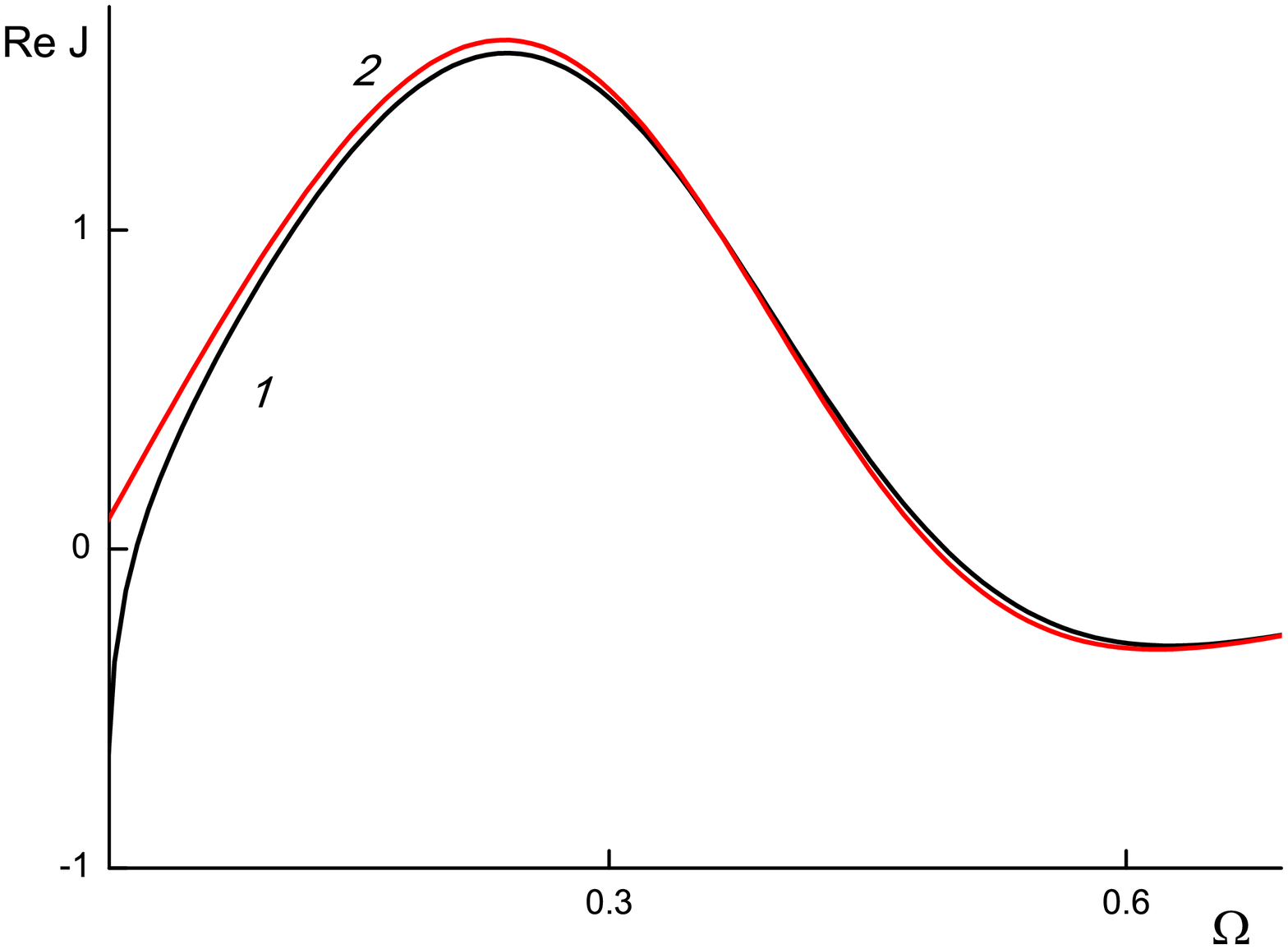}
{{\bf Fig. 5.}  Real part of density of dimensionless current, $q=0.3$.}
\includegraphics[width=16.0cm, height=10cm]{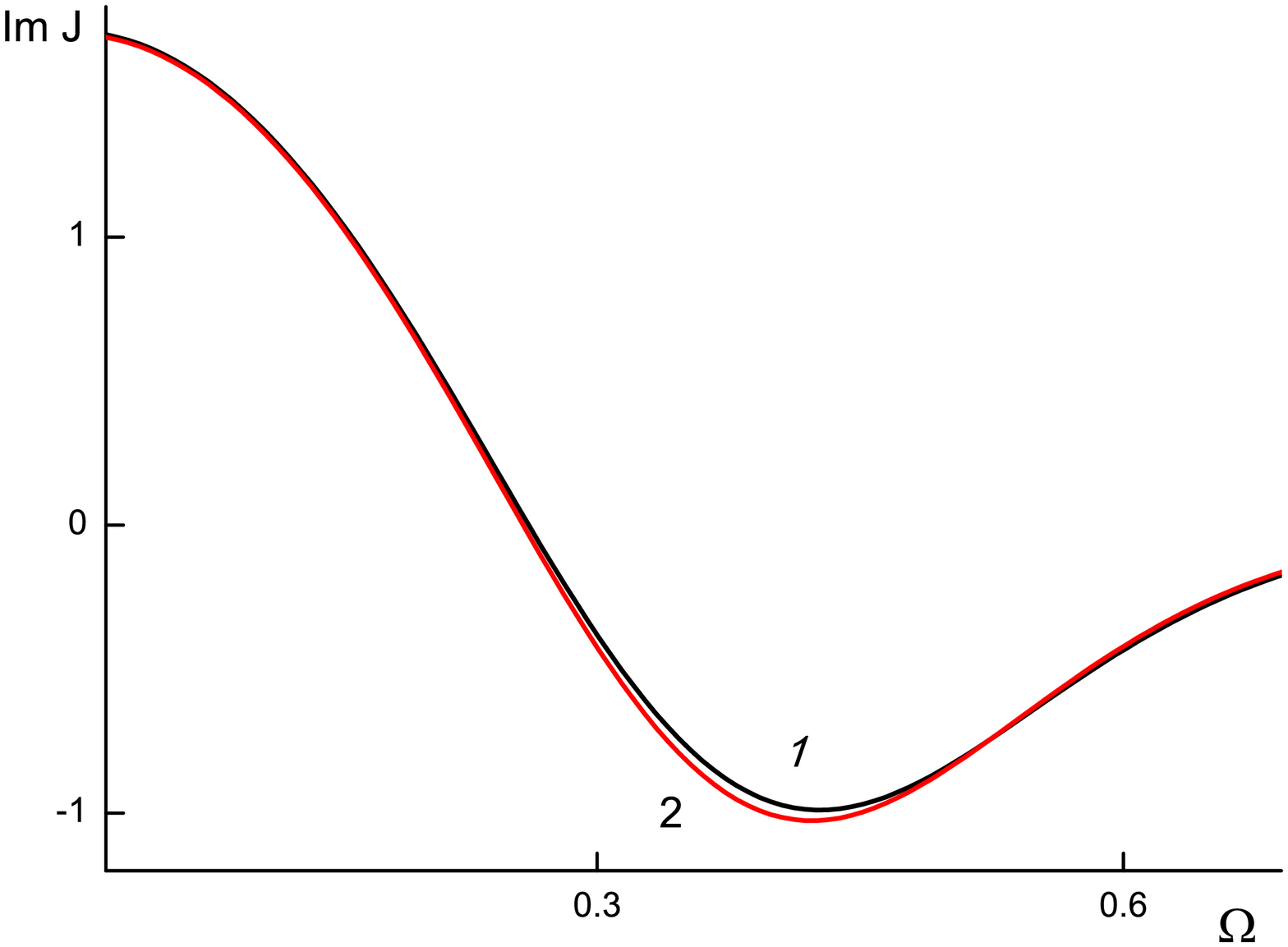}
{{\bf Fig. 6.} Imaginary part of density of dimensionless current, $q=0.3$.}
\end{figure}

\begin{figure}[thp]\center
\includegraphics[width=16.0cm, height=10cm]{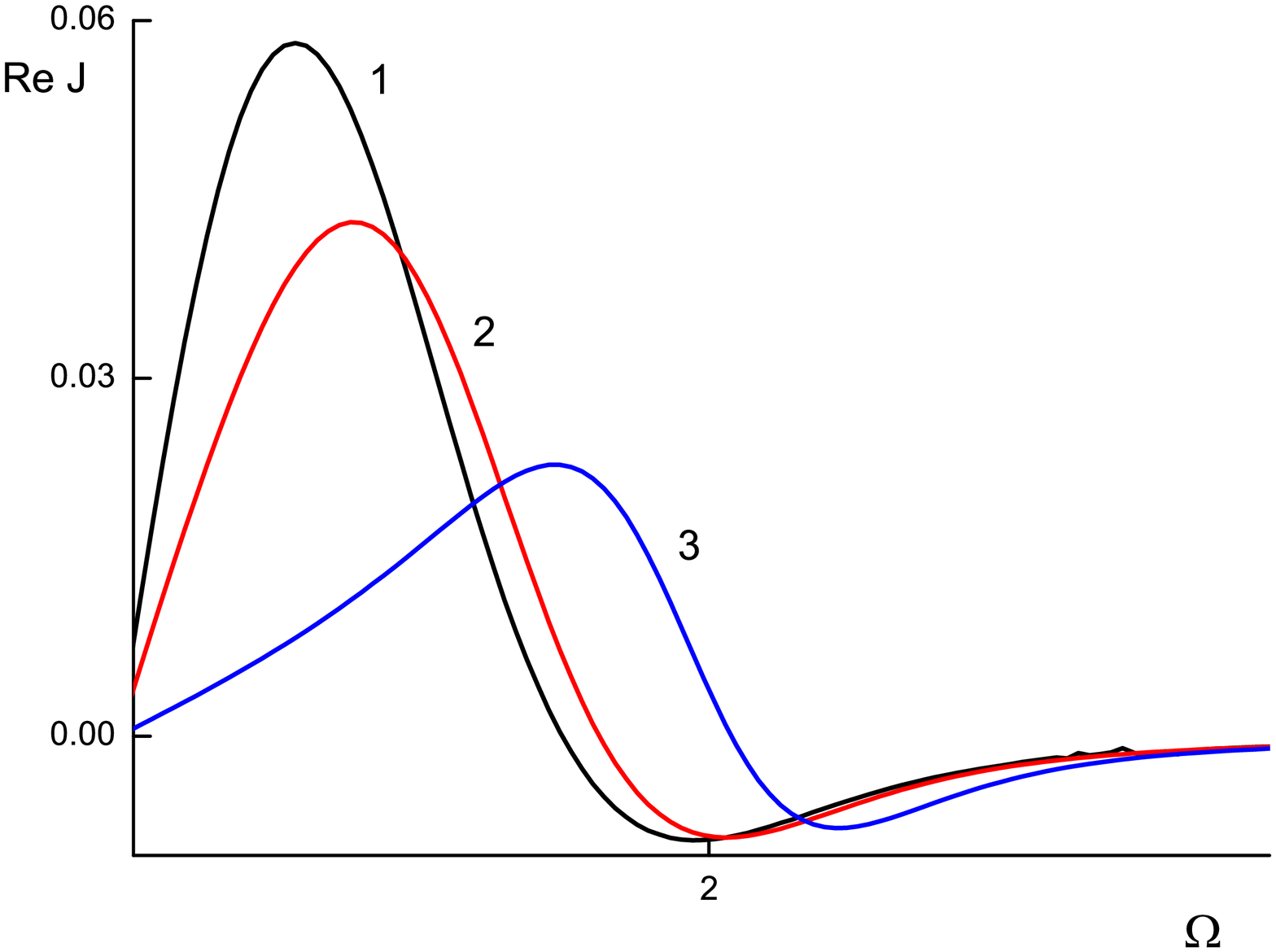}
{{\bf Fig. 7.}  Real part of density of dimensionless current
in classical plasma, $q=1$.
Curves 1,2,3 correspond to values of dimensionless chemical potential
$\alpha=-5,0,3$.}
\includegraphics[width=16.0cm, height=10cm]{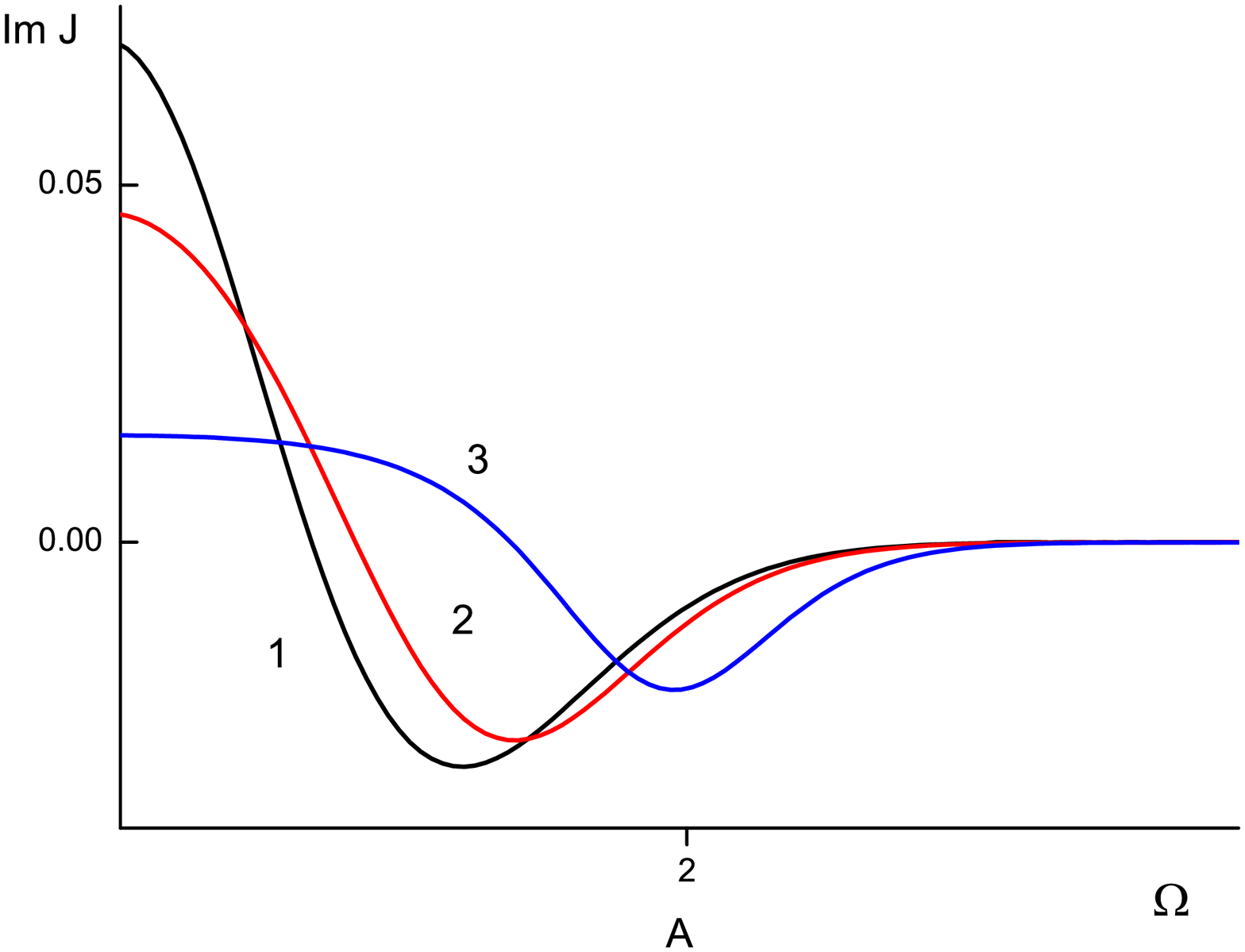}
{{\bf Fig. 8.} Imaginary part of density of dimensionless current
in classical plasma, $q=1$.
Curves 1,2,3 correspond to values of dimensionless chemical potential
$\alpha=-5,0,3$.}
\end{figure}

\begin{figure}[thp]\center
\includegraphics[width=16.0cm, height=10cm]{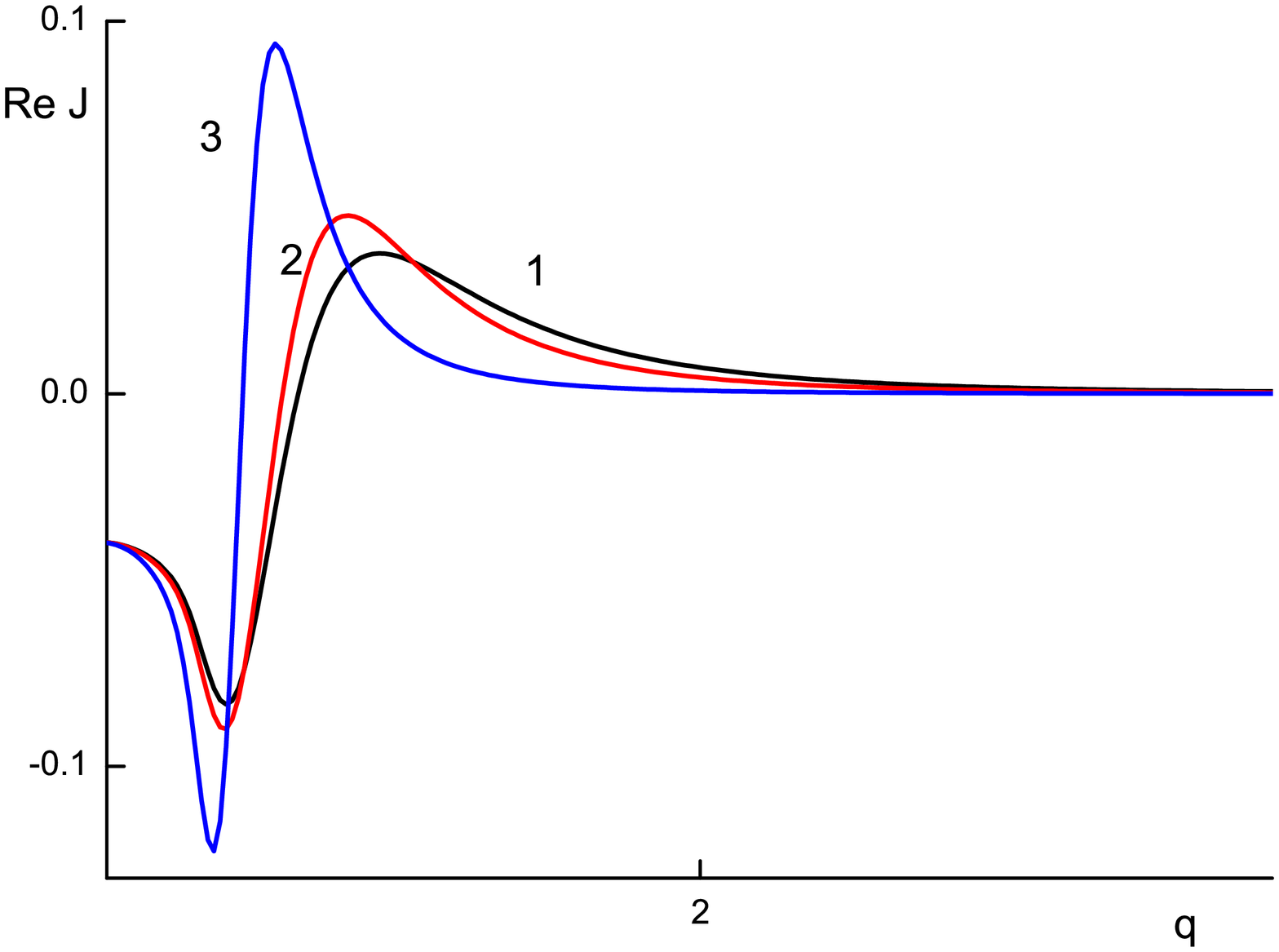}
{{\bf Fig. 9.} Real part of density of dimensionless current
in classical plasma, $\Omega=1$.
Curves 1,2,3 correspond to values of dimensionless chemical potential
$\alpha=-5,0,3$.}
\includegraphics[width=16.0cm, height=10cm]{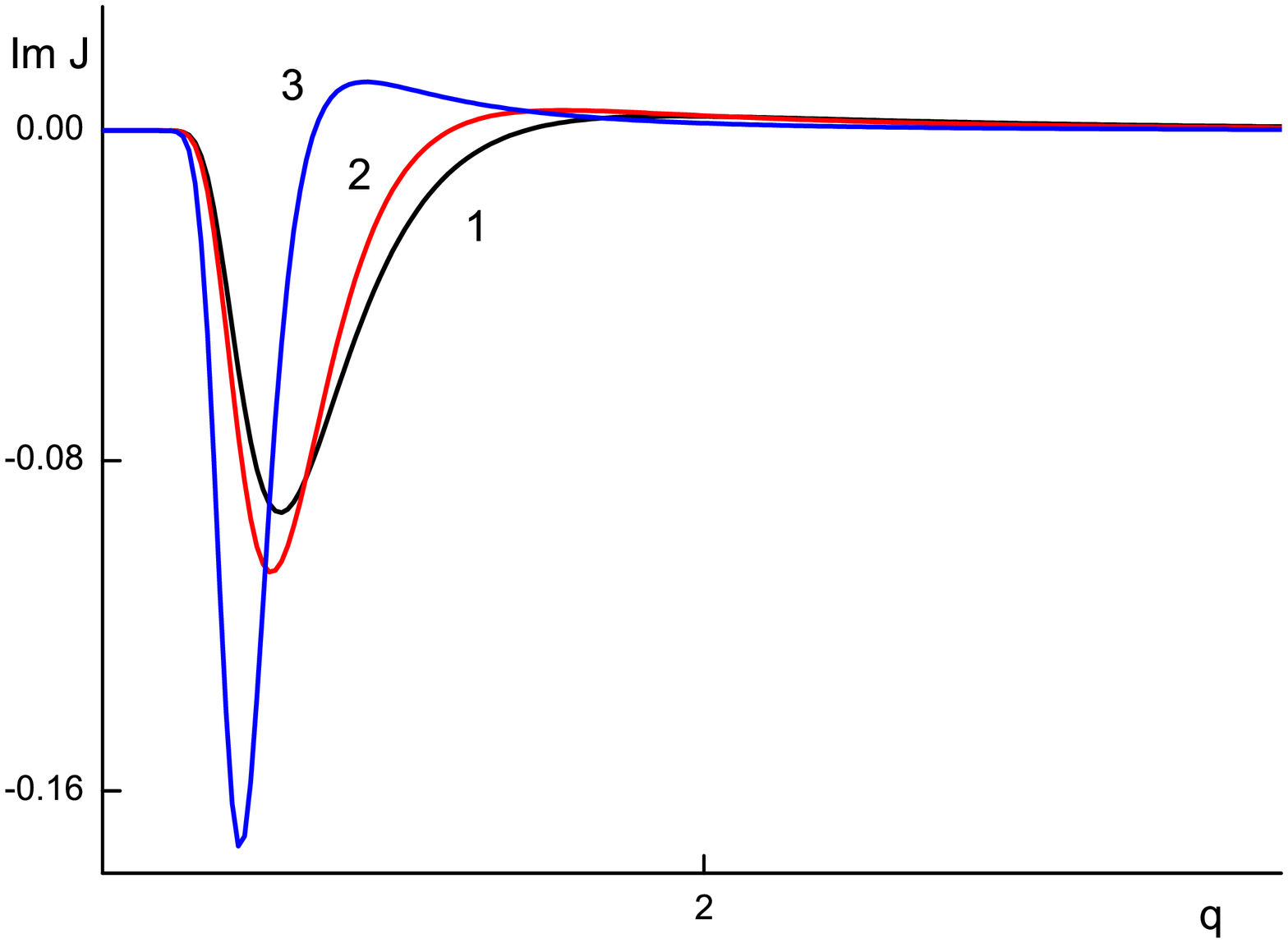}
{{\bf Fig. 10.}  Imaginary part of density of dimensionless current
in classical plasma, $\Omega=1$.
Curves 1,2,3 correspond to values of dimensionless chemical potential
$\alpha=-5,0,3$.}
\end{figure}

\clearpage

\end{document}